\documentclass[twocolumn,showpacs,preprintnumbers,amsmath,amssymb,floatfix]{revtex4}
\usepackage{subfigure}
\usepackage{amsmath}
\usepackage{amsfonts}
\usepackage{amssymb}
\usepackage[dvips]{graphicx}

\usepackage[colorlinks]{hyperref}
\usepackage[usenames,dvipsnames]{color}

\usepackage{IEEEtrantools}
\usepackage[applemac]{inputenc}
\usepackage[T1]{fontenc} 
\usepackage{amsmath} 
\usepackage{amsfonts} 
\usepackage{amssymb}
\usepackage{graphicx} 

\newcommand{\half}{\tfrac{1}{2}}
\def\beq{\begin{equation}}
\def\eeq{\end{equation}}
\def\bea{\begin{eqnarray}}
\def\eea{\end{eqnarray}}
\def\ba{\begin{array}}
\def\ea{\end{array}}

\def\lb{\left(}
\def\rb{\right)}

\def\l.{\left.}
\def\r.{\right.}

\def\ie{{\it i.e. }}

\def\ame{&=&}

\def\part{\partial}
\def\tfrac#1#2{{\textstyle{#1\over #2}}}
\def\half{\tfrac{1}{2}}

\def\nne{\nonumber\\}

\begin{document}
\preprint{UdeM-GPP-TH-18-263}
\title{The twin paradox: the role of acceleration}
\author{J. Gamboa$^{1}$} 
\email{jorge.gamboa@usach.cl}
\author{F. Mendez$^{1}$} 
\email{fernando.mendez@usach.cl}
\author{M. B. Paranjape$^{1,2}$} 
\email{paranj@lps.umontreal.ca}
\author{Benoit Sirois$^{2}$} 
\email{benoit.sirois@umontreal.ca}
\affiliation{$^{1}$Departamento de F\'\i sica, Universidad de Santiago de Chile, Avenida Ecuador 3493, Estación Central, Santiago, Chile 9170124.}
\affiliation{$^{2}$Groupe de physique des particules, D\'epartement de physique,
Universit\'e de Montr\'eal,
C.P. 6128, succ. centre-ville, Montr\'eal, 
Qu\'ebec, Canada, H3C 3J7 }

\begin{abstract} 
The twin paradox, which evokes from the the idea that two twins may age differently because of their relative motion, has been studied and explained ever since it was first described in 1906, the year after special relativity was invented.  The question can be asked:  ``Is there anything more to say?''  It seems evident that acceleration has a role to play, however this role has largely been brushed aside since it is not required in calculating, in a preferred reference frame, the relative age difference of the twins.  Indeed, if one tries to calculate the age difference from the point of the view of the twin that undergoes the acceleration, then the role of the acceleration is crucial and cannot be dismissed.  In the resolution of the twin paradox, the role of the acceleration has been denigrated to the extent that it has been treated as a red-herring.  This is a mistake and shows a clear misunderstanding of the twin paradox.  
\end{abstract}

\pacs{~}

\maketitle

\section{Introduction}

The twin paradox corresponds to the following set of events and observations.  Two twins are at rest in an inertial reference frame.  Both carry physical clocks that are synchronized.  One of the twins stays put, we will call this twin A.  The other twin, who we will call B, takes a trip and goes away for a while and then returns, usually at relativistic speeds to bring out the paradox.  When the twins are back together, they compare their clocks and they find that the clock of  B  shows less time to have elapsed than the clock of  A, thus  B is younger.  

The paradox exists at two levels.  The first level is for those unfamiliar with special relativity, for this cohort, everyone ages at the same rate.  Hence they ask ``How can one twin become younger than the other?''  We can dispense with the paradox for this cohort by simply exhorting them to go learn what special relativity has to say.  We will not consider it further.  

At the second level, for the cohort familiar with special relativity, it is clear that because of the motion, due to special relativistic time dilation, the clock of  B must be slower.  However, the paradox returns for a specific journey as shown in Fig. \eqref{fig1}, in the specific approximation where we neglect the accelerating parts of the journey of B.  Consider the journey where A remains always at rest, but B first accelerates forward for a short time, then coasts at constant velocity for a long time, then decelerates to velocity zero and continues to accelerates backwards for a short time until  his or her velocity has reversed, then coasts again for a long time and finally accelerates forward for a short time until his or her velocity is zero and he or she is at rest beside  A.   If the accelerating periods can be taken to be very short compared to the coasting periods then we can imagine the approximation that they can be neglected is valid.  Then differential aging can be calculated simply from the periods of coasting, where special relativity is valid.  Now the paradox appears due to the apparent symmetry of the situation.  During the coasting periods as special relativity is valid, all motion is strictly relative.  Hence it is equally valid that A thinks B has moved and then comes back as it is for B to think A has moved in the opposite direction and then comes back.  If we can neglect the periods of acceleration, the symmetry of the situation yields the paradox that each twin thinks the other twin must be younger.  Of course this analysis is fallacious.

What is the crucial difference between the two twins that invalidates this analysis and eventually resolves the paradox?  It is the fact that one of the twins has not always been in an inertial reference frame, B has suffered acceleration.   The point is, that even if one neglects the actual periods of acceleration, the motion of B is simply not symmetric with respect to the motion of A.  There is a physical turn around point for B, say a star to which B goes to and then comes back from, and importantly, the turn around point is at rest relative to A but not at rest relative to B.  

The distance to this turn around point is strictly not the same in the inertial coordinate system of A and that of B.   Initially, when the coasting begins, if A sees the turn around point at a distance  $X_F$ away, then A sees B coasting towards the turn around point for a time $X_F/V$, where $V$ is the coasting velocity.  On the other hand from B's point of view, B sees the turn around point approaching with  (coasting) velocity $-V$.  Then the simple Lorentz contraction gives that B finds that the (initial) distance to the turn around point is only $X_F\sqrt{1-V^2}$.  Correspondingly, B observes A receding (coasting) for a time $X_F\sqrt{1-V^2}/V$.   This amount of elapsed (coordinate) time is clearly less than  $X_F/V$, the time A observes B coasting to the turn around point. Therefore, the coasting periods associated with each twin are simply not symmetric.  This is explicitly and solely because it is twin B who actually moves, twin B physically accelerates and turns around.  It is the goal of this tutorial to elucidate precisely and quantitatively  the role of this acceleration in the twin paradox.

The original twin paradox being as old as relativity itself, has obviously been studied extensively since its formulation. Many analyses exist to date.  Some analyses study light signals and their arrival times, sent between the twins \cite{Atrip}.  Some classical analyses perform a full calculation of the elapsed time from the point of view of both twins, in the case of smooth acceleration periods \cite{Moller,Perrin}, or in the limit of instantaneous accelerations \cite{Muller}.  However, these analyses thoroughly use the ideas and methods of general relativity and in that way although closest in spirit, are not close in content to the analysis that we will be present here.  We will distinguish our calculations from the previous ones by exemplifying the crucial role of acceleration.  The physical implications of these calculations has been discussed in \cite{Gbuilder}.  Introductory relativity discussions of the twin paradox typically avoid analyzing the full version of the paradox which must include the role of acceleration.   It is held, (wrongly, see for example \cite{Uniformlyaccelerated}), that the study of accelerated reference frames requires prior knowledge of general relativity.  We find this to be a pity.  

It is clear that the calculation of the lapse of proper time for each twin can be done in any reference frame.  The lapse of proper time being an invariant under change of coordinates, is independent of the frame in which it is calculated.  With this understanding, there is of course no twin paradox.  However, we still find it instructive to be able to explicitly compute the elapsed proper time for each twin, according to each twin.  The paradox specifically reappears because of the idea, which is false, that it is valid to neglect the acceleration and the fact that we know which is the twin that takes the journey because of the acceleration, in the calculation of the relative elapsed proper time, in any frame.    Indeed, under the assumption of acceleration for short times compared to the coast times, it is perfectly correct to compute the relative aging, using only one's knowledge of special relativity, and neglecting the accelerating parts of the journey, but only in the reference frame of the twin that does not take a trip.  It is on the other hand, completely false to believe that this same neglect is correct in every other reference frame, specifically in the reference frame of the twin that does actually take the journey.   For this, accelerated, moving twin, it is crucial to take into account the acceleration.  From the point of view of the accelerated twin, most of the aging of the unaccelerated twin actually occurs during the accelerating phase.

Consequently, the aim of this tutorial is twofold: firstly, we wish to present a calculation of the lapse of proper time of each twin, according to each twin, using only change of variables and without any recourse to the machinery of general relativity and differential geometry as done in \cite{Perrin}.  This will require, to bring home the point, the derivation of the equations of motions of the movement of A, as perceived by B, which is simpler and hopefully clearer than the accounts already found in literature \cite{Moller,Perrin, Muller}. These accounts are sometimes quite old, they carry many notational conventions which are nowadays obsolete, and which make their reading very dense for anybody, especially students who are not already well accustomed with relativity. Secondly, we would like to discuss in which respect this calculation helps in understanding the twin paradox. Specifically, we will comment on many points recently presented by Maudlin in his latest book \textit{Philosophy of Physics: Space and Time} \cite{Maudlin2012}, especially  his position on the non-importance of acceleration of B.

\section{The twin paradox}

Let us start by stating exactly which version of the twin paradox will be analyzed here: two twins live in a flat, infinite, 1+1 dimensional Minkowski spacetime. One twin will take a journey through space, to a fixed point in Minkowski space, and then return, eventually coming back to his or her sibling. The twin who remains at rest the whole time will be referred to as A, while the twin who takes the journey will be referred to as B.  Twin A is at rest in the given Minkowski space time, while twin B moves through it, accelerating at times.  It is the neglect of this asymmetry between the two twins that gives rise to the twin paradox, and taking the asymmetry into account resolves it.  Just before B begins his or her journey, they both synchronize their clocks, such that both clocks indicate 0.  B accelerates, then coasts for some time, goes through another acceleration period in which he or she reverses his or her direction and velocity, and then coasts back towards A to finally decelerate one last time, stopping exactly so that he or she arrives at rest next to A.  All 4 periods of acceleration (velocity of B in the A's frame varies as: $0 \rightarrow V; V \rightarrow 0; 0 \rightarrow -V; -V \rightarrow 0$) are symmetrical and  characterized by a constant proper acceleration of value $g$. Both coasting periods are also somewhat symmetrical. When B stops, the twins compare their clocks, and we will show that B will be younger than A.  We will calculate the total elapsed proper time during the trip for each twin, first in the frame of reference of A and then in the frame of reference of B. 

\section{Elapsed proper time, for and by each twin} \label{sec3}
In this section we will explicitly calculate the elapsed proper time of each twin, and we will do the calculation twice, once according to each twin.  For convenience, we will use units in which the speed of light is unity, $c=1$.  We will denote the elapsed proper time of each twin by $\Delta\tau_{\rm A}(\rm A)$ and  $\Delta\tau_{\rm B}(\rm A)$ for A and B respectively as calculated by A, and correspondingly, we will denote elapsed proper time of each twin by $\Delta\tau_{\rm A}(\rm B)$  and  $\Delta\tau_{\rm B}(\rm B)$ for A and  B respectively but this time as calculated by B.  We will show explicitly that these are in fact independent of which twin does the calculation, that is we will show:
\bea
\Delta\tau_{\rm A}(\rm A)\ame\Delta\tau_{\rm A}(\rm B)\\
\Delta\tau_{\rm B}(\rm A)\ame\Delta\tau_{\rm B}(\rm B)
\eea
The proper time calculated in an inertial reference frame is Lorentz invariant.  This is because the metric of Minkowski spacetime, which defines the proper time, is Lorentz invariant
\beq
d\tau^2=dT^2-dX^2.\label{mm}
\eeq
 Twin A is of course always in an inertial reference frame hence invariance under Lorentz transformations is expected.  However, Twin B  is not.  Therefore, the calculation of the proper times according to twin B will come out to be equal to those calculated by twin A because in fact the proper time is not only Lorentz invariant, but is diffeomorphism invariant, a fancy name for invariant under an arbitrary change of coordinates, \ie it is invariant for absolutely any observer.   The analysis of this arbitrary diffeomorphism invariance is out of the scope of the present article and will not be presented here.

It will become clear that it is much easier to do the calculation in the reference frame of A, however, it is also necessary to explicitly do the calculation in the reference frame of B to show that indeed B will also find exactly the same values for the elapsed proper times and hence, once and for all,  dispel with the paradox.  Most importantly, for the calculation according to B, it will be seen that it is crucial not to neglect the periods of acceleration.  We will refer to coordinates according to A by the notation  $x^\mu = (T,X)$ while the coordinates according to B will be denoted as  $\tilde x^\mu = (\tilde T,\tilde X)$. 
\subsection{Elapsed proper time of A  according to A}
The Lorentz frame in which A is always at rest, hereby referred to as $R$, is equipped with coordinates $x^\mu = (T,X)$. In this frame, the lapse of proper time of A between events at fixed $X$ is simply equal to the lapse of coordinate time $T$.  Setting $c = 1$, with a metric signature $(+,-)$, we have the infinitesimal elapsed proper time, as in Eqn.\eqref{mm}, for infinitesimal elapsed coordinate time $dT$ and infinitesimal change of spatial coordinate $dX$
\begin{equation} \label{eq1}
\mathrm{d}\tau^2(\rm A) = \mathrm{d}T^2 - \mathrm{d}X^2\,
\end{equation}
where the notation $\mathrm{d}\tau(\rm A)$ indicates the proper time according to A.  $dX=0$ when A is at rest, therefore, $(T,X)=(T_{\rm  A},0)$ for the trajectory of A in A's coordinates, and hence
\beq
d\tau_{\rm A}({\rm  A})=dT_{\rm A}\label{propertimeA}
\eeq
where now the notation $d\tau_{\rm A}({\rm  A})$ indicates the proper time of A according to A, and also  treating $\tau_{\rm A}({\rm A})$ as a function of the coordinate time of A, $T_{\rm A}$.   Thus we see that the elapsed proper time for A is equal to the elapsed coordinate time for A.  Then we have
\beq
\Delta\tau_{\rm A}({\rm A})=\int_0^{\Delta\tau_{\rm A}({\rm A})}d\tau_{\rm A}({\rm  A})= \int_0^{T_F}dT_{\rm A}=T_F
\eeq
where $T_F$ is defined as the elapsed coordinate time for A when B has returned (and from the calculation, we see that it is, equally well, numerically equal to the elapsed proper time for A).  

\begin{figure}
\begin{center}
\includegraphics[scale=0.5]{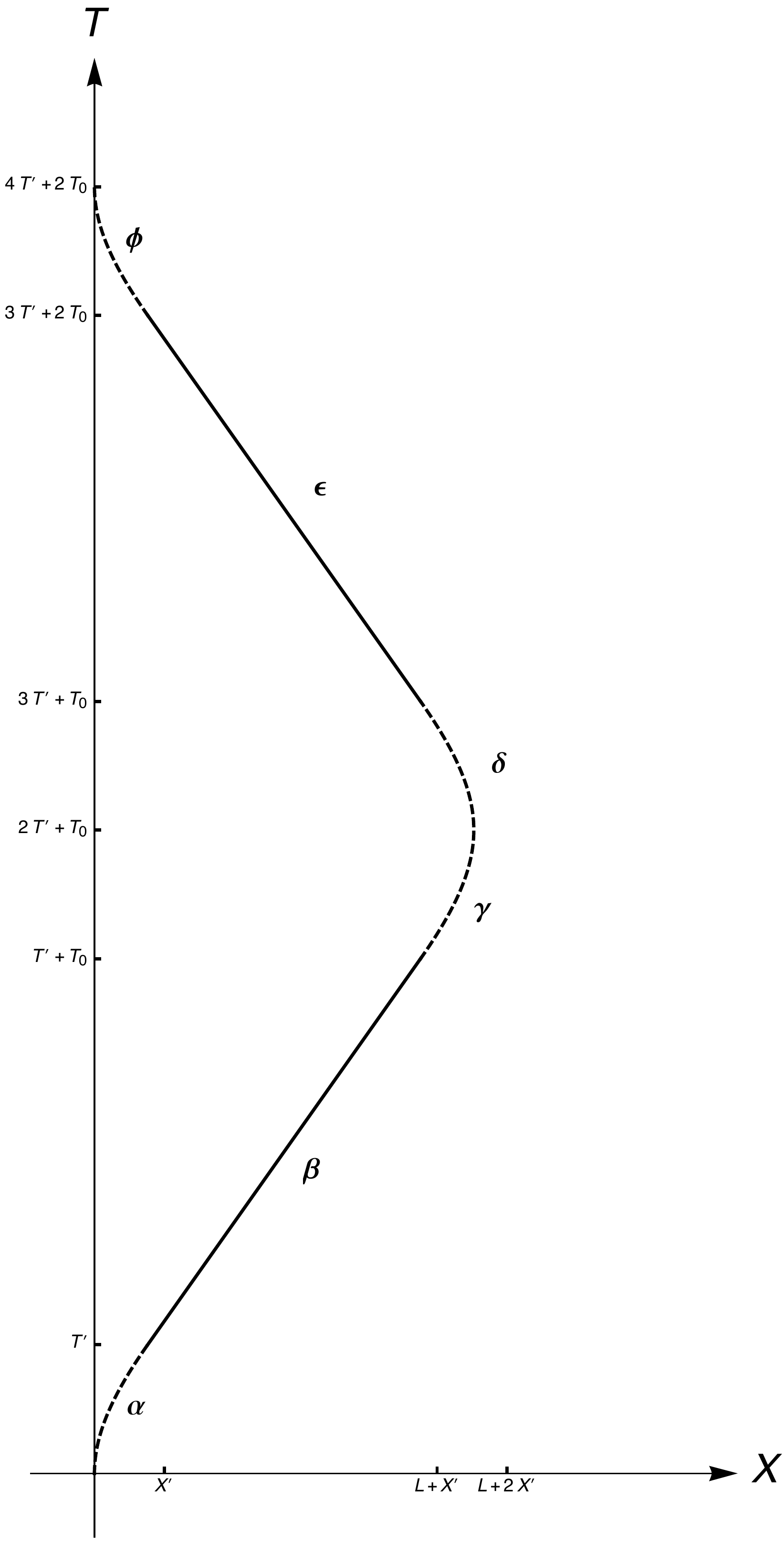}
\caption{Worldline of B, illustrated in reference frame $R$}\label{fig1}
\end{center}
\end{figure}

We will find it useful to express $T_F$ in terms of some intermediate times that are relevant to the motion of B, as seen by A.  Figure \eqref{fig1} illustrates the path of B through spacetime as seen in reference frame $R$, the dashed segments representing the accelerating periods and the solid segments the coasting periods. The greek letters label different episodes in the motion of B.  The first accelerating period of B is referred to as  $\alpha$, the first coasting period as $\beta$, the second and third accelerating periods as $\gamma$ and $\delta$, the second coasting period as $\epsilon$ and the final accelerating period as $\phi$.  It should be evident that A's worldline  simply lies along the $T$ axis. 

$(T_{\rm B},X_{\rm B})$ are the coordinates of B according to A on its worldline in Figure\eqref{fig1}.  $(T',X')$ are the coordinates of B, according to A, when the first period of acceleration $\alpha$ ends.  A little reflection will convince the reader that each accelerating period of B lasts for the same amount of time $T'$ in the reference frame $R$.  

$L$ is the spatial length and $T_0$ is the elapsed coordinate time in the reference frame $R$, of the coasting periods of B,  which are labelled $\beta$ and $\epsilon$. 
In reference frame $R$, B has velocity $V$ and $-V$ during the coasting periods and clearly $V=L/T_0$.  The world line of B, for the coasting period $\beta$ for example, with coordinates $(T_{\rm  B},X_{\rm  B})$ is simply found by writing down the equation of a straight line with the correct slope and then ensuring that it passes through the point $(T',X')$ as
\bea
T_{\rm B}&=&\frac{(T_0+T')-T'}{(L+X')-X'}X_{\rm B} +T' - \frac{(T_0+T')-T'}{(L+X')-X'}X'\nonumber\\
&=&\frac{T_0}{L}X_{\rm B}+T'-\frac{T_0}{L}X'\nonumber\\
&=&\frac{X_{\rm B}}{V}+T'-\frac{X'}{V}.\label{eqofB}
\eea
We reiterate, $T_{\rm B}$ and $X_{\rm B}$ are the instantaneous time and spatial coordinates of B,  $T'$ and  $X'$ the coordinates when the first acceleration period ends and $T_0$ and $L$ are the elapsed coordinate time and distance respectively during which B is coasting, and all coordinates given in the reference frame $R$ (\ie all according to A) and are denoted on Figure \eqref{fig1}.
Then the elapsed proper time of A, during the coasting periods of B according to A, integrating Eqn.\eqref{propertimeA} is
\begin{equation} 
\tau_{{\rm A},\beta}({\rm A}) = \tau_{{\rm A},\epsilon}({\rm A}) \equiv T_0 = \frac{L}{V}\, .
\end{equation}
\noindent

During all of the accelerating periods the amount of time that will pass on A's clock is always the same and denoted by $T'$.  Then the elapsed proper time is also given by $T'$,
\beq
\tau_{{\rm A},\alpha}({\rm A})=\tau_{{\rm A},\gamma}({\rm A})=\tau_{{\rm A},\delta}({\rm A})=\tau_{{\rm A},\phi}({\rm A})=T'.
\eeq
 Therefore, the total elapsed proper time $\Delta\tau_{\rm A}({\rm A})=T_F$ for A, making references to certain time stamps that A makes for the trip of B, for the entire round trip of B  is equal to
\begin{equation}
\Delta\tau_{\rm A}({\rm A})=T_F= 2T_0 + 4T'\,.\label{7}
\end{equation}
\subsection{Elapsed proper time of B  according to A}
The elapsed proper time for B according to A requires two calculations, one for the coasting periods and one for the accelerating periods.  Clearly, by symmetry, the elapsed proper time of B according to A is the same in the two coasting periods and separately the same in the four accelerating periods.  The motion of B for the period $\beta$ where $V$ is constant, satisfies, inverting Eqn.\eqref{eqofB}
\beq
X_{\rm B}=VT_{\rm B}-VT'+X'.
\eeq
Then the differential elapsed proper time of B according to A satisfies
\bea
\mathrm{d}\tau^2_{\rm B}({\rm A})&=&dT_{\rm B}^2-dX_{\rm B}^2\nne
\ame \lb1-\lb\frac{dX_{\rm B}}{dT_{\rm B}}\rb^2\rb dT_{\rm B}^2\nne
\ame \lb1-V^2\rb dT_{\rm B}^2
\eea
and hence
\begin{IEEEeqnarray}{rCl}
\tau_{{\rm B},\beta}({\rm A}) 
& = &  \int_{T'}^{T_0+T'}\left(1 - V^2\right)^{\frac{1}{2}} \mathrm{d}T_{\rm B} \nonumber\\
& = & \left(1 - V^2\right)^{\frac{1}{2}} T_0\nonumber\\
\label{eq4}
& = & \frac{T_0}{\gamma_V }= \tau_{{\rm B},\epsilon}({\rm A})
\end{IEEEeqnarray}
where  $\gamma_V=1/\sqrt{1-V^2}$ is the standard Lorentz factor, and once and for all $\tau_{\rm B}({\rm A})$ is the elapsed proper time of B according to A (we expect that the notation is now clear and we will not have to explain it each time).  $\gamma_V$ could be a large number, making in principle,  that the proper time that elapses, during the coasting periods for B according to A,  is very much smaller than the elapsed proper time for A during these periods.  The standard presentation of the twin paradox is based on this understanding, that A ages $T_0$ during the coasting phase of B, while B ages $T_0/\gamma_V<T_0$, hence B is younger than A.  Because B is in motion according to A, its elapsed proper time is smaller than the elapsed proper time of A, for the same elapsed coordinate time.  

As this is during the coasting period when both twins are in inertial reference frames, exactly the same analysis would apply for B concerning A, and then we are in a paradoxical situation where each twin thinks the other must be younger.  The resolution of the paradox comes through the understanding that the two periods are not symmetrical and additionally through examining what happens during the accelerating periods.  

For  all the accelerating periods, we can first observe that each accelerating period will last the exactly same amount of time on A's clock. This is due to symmetry under time reversal and parity (invariance under spatial reflexion).  According to A, B accelerating from 0 to $V$, in period $\alpha$, will take the same amount of coordinate time as B accelerating from $-V$ to 0 in period $\phi$ due to time reversal invariance.  Then by parity this amount of coordinate time is equal to the coordinate time it will take to decelerate  from $V$ to 0, which is period $\gamma$. Decelerating from 0 to $-V$, in period $\delta$, will take the same amount of coordinate time as accelerating from 0 to $V$ of period $\alpha$, due to invariance under parity.    We will therefore choose to analyze period $\alpha$ which will be simplest.  First we must understand acceleration in the context of special relativity. 

\subsubsection{Acceleration in special relativity}
We imagine an object moving with velocity $V(T)$ in the reference frame  $R$.   Then the object's acceleration is defined in the Newtonian way\footnote{Be careful note to confuse the acceleration $A(T)$ (font math italics) with the label for the twin that stays put, A (font roman)}
\beq
A(T)=\frac{dV(T)}{dT}.
\eeq
This is clearly the acceleration that an observer in reference frame $R$ would measure.  How is this acceleration perceived by an observer moving with the object?  This is not a perfectly ``special relativity'' question, since if the object is continually accelerating, then so is the observer, and therefore this observer is not always an inertial observer.  However, we can imagine that at each time, there is an inertial observer that passes by our object with exactly the instantaneous velocity of our object, and we can ask what is the acceleration that is perceived by such an observer at each time.  This is a well defined question that is within the purview of special relativity.  We must transform to the coordinates of the observer moving with the instantaneous velocity of the object.  

For simplicity, we will analyze acceleration in only one direction, say the $x$ direction.  We consider an object moving arbitrarily in reference frame $R$ with trajectory given by the coordinates $(X(T),T)$.   Then in a reference frame $\bar R$, which is moving at a fixed (time independent) velocity $v$ relative to $R$, the motion of the object will have the trajectory $(\bar X(\bar T),\bar T)$.  The standard formula for the (Lorentz) transformation of the coordinates, at each given moment, is given by:
\bea
\bar X&=&\gamma_v (X-vT)\label{Xprime}\\
\bar T&=&\gamma_v (T-v X)\label{Tprime}
\eea
where $\gamma_v=1/\sqrt{1-v^2}$.   
We will also need the inverse relations:
\bea
X&=&\gamma_v (\bar X+v\bar T)\label{X}\\
T&=&\gamma_v (\bar T+v \bar X)\label{T}
\eea
We emphasize that $v$ is the constant, time independent, relative velocity between the two reference frames $R$ and $\bar R$.  It should not be confused with what we will call $V$ or $\bar V$, which are the velocity of the object moving with an arbitrary trajectory as seen by an observer in each reference frame, respectively.  
The velocity of the object in reference frame $R$ is simply
\beq
V(T)=\frac{dX(T)}{dT}
\eeq
and the velocity  of the object in reference frame $\bar R$ is
\beq
\bar V(\bar T)=\frac{d\bar X(\bar T)}{d\bar T},
\eeq
while the acceleration of the object in reference frame $R$ is simply
\beq
A(T)=\frac{d^2X(T)}{dT^2}=\frac{dV(T)}{dT}
\eeq
and the acceleration  of the object in reference frame $\bar R$ is
\beq
\bar A(\bar T)=\frac{d\bar X^2(\bar T)}{d\bar T^2}=\frac{d\bar V(\bar T)}{d\bar T}.
\eeq
To find the formula for the transformation of acceleration we must differentiate Eqn.\eqref{Xprime} twice with respect to $\bar T$, and then express the result in terms of $v$, $V(T)$ and $A(T)$.  Differentiating once we find
\bea
\bar V(\bar T)&=&\frac{d\bar X(\bar T)}{d\bar T}=\gamma_v\left(\frac{dX(T)}{d\bar T}-v\frac{dT}{d\bar T}\right)\nne
&=&\gamma_v\left(\frac{dX(T)}{dT}-v\right)\frac{dT}{d\bar T}=\gamma_v\left(V(T)-v\right)\frac{dT}{d\bar T}.\nne
\eea
Now differentiating Eqn.\eqref{T} with respect to $\bar T$ we get
\beq
\frac{dT}{d\bar T}=\gamma_v\left(1+v\bar V(\bar T)\right)
\eeq
therefore we get
\beq
\bar V(\bar T)=\gamma_v \left( V(T)-v\right)\gamma_v \left( 1+v\bar V(\bar T)\right).
\eeq
This linear relation for $\bar V(\bar T)$ can easily be solved as
\beq
\bar V(\bar T)=\lb\frac{V(T)-v}{1-vV(T)}\rb\label{Vprime}
\eeq
which is the well known formula for the addition of velocities.  Taking another derivative with respect to $\bar T$ gives
\bea
\bar A(\bar T)\ame\frac{d\bar V(\bar T)}{d\bar T}\nne
&=&\frac{1}{(1-vV(T))}\frac{dV(T)}{d\bar T}-\frac{(V(T)-v)}{(1-vV(T))^2}(-v)\frac{dV(T)}{d\bar T}\nne
&=&\frac{(1-v^2)}{(1-vV(T))^2}\frac{dV(T)}{dT}\frac{dT}{d\bar T}\nne
&=&\frac{(1-v^2)}{(1-vV(T))^2} A(T)\gamma_v\lb 1-v\bar V(\bar T)\rb .
\eea
Replacing for $\bar V$ from Eqn.\eqref{Vprime} gives the formula for the transformation of the acceleration
\beq
\bar A(\bar T)=\frac{A(T)}{\gamma_v^3\lb 1-vV(T)\rb^3}.\label{af}
\eeq
\subsubsection{The uniformly accelerated observer}
The notion of uniform acceleration makes sense in the following way.  At each instant during the acceleration, the inertial observer travelling at the instantaneous velocity of the accelerated object will see the accelerated object as moving non-relativistically and will actually be accelerating from rest (of course the inertial observer could have a small relative velocity will also be fine for this analysis, but the analysis is clearest with zero relative velocity).  Hence we can be confident that the acceleration this observer measures is given by the non-relativistic formula.   Then we can impose that such observers (a different one at each instant) always measure the same acceleration, as the definition of a uniformly accelerating object.  

Therefore we impose, for each observer (different observers at different times) travelling with velocity $v=V(T)$, the acceleration that each one measures is a constant that we call $g$, using Eqn.\eqref{af}:
\bea
\bar A(\bar T)|_{v=V(T)}&\equiv& g=\frac{A(T)}{\gamma_{V(T)}^3\lb 1-(V(T))^2\rb^3}=\gamma_{V(T)} ^3A(T)\nne
\ame\gamma_{V(T)}^3\frac{dV(T)}{dT}.
\eea
This yields the differential equation for $V(T)$
\beq
\frac{dV(T)}{dT}=g(1-V^2(T))^{\frac{3}{2}}\label{eq5}
\eeq
which, with the initial condition that $V=0$ at $T=0$,  integrates as
\beq
V(T)= \frac{gT}{\left(1+(gT)^{2}\right)^{\frac{1}{2}}}.\label{uaccv}
\eeq
This formula is easily inverted as
\beq
T(V)=\frac{1}{g}\frac{V(T)}{\sqrt{1-(V(T))^2}}.
\eeq
Therefore we can identify $T'$, the coordinate time of twin A at the end of the acceleration as seen in Fig.\eqref{fig1}, as a function of the coasting velocity $V$ as
\beq
T'=\frac{1}{g}\frac{V}{\sqrt{1-V^2}}\label{tpv}
\eeq
and equally well
\beq
V=V(T')= \frac{gT'}{\left(1+(gT')^{2}\right)^{\frac{1}{2}}}.
\eeq
We can integrate the formula of Eqn.\eqref{uaccv} once to find the position of the uniformly accelerated object as a function of coordinate time as
\beq
X(T)=\frac{1}{g}\lb\lb1+\lb gT\rb^2\rb^{\half}-1\rb +X(0).
\eeq

\subsubsection{Proper time of {\rm B} as calculated by  {\rm A} }
We can now use this formula of Eqn.\eqref{uaccv} for the velocity of a uniformly accelerated object as a function of the coordinate time $T$, to find the proper time of B as a function of its coordinate time $T_{\rm B}$ according to A.  If we write the (differential) of the proper time of B, using Eqn. (\ref{eq1}),  in terms of the coordinates of B  according to A, we find
\bea
\mathrm{d}\tau^2_{\rm B}({\rm A})&=&dT_{\rm B}^2-dX_{\rm B}^2\label{eq28}\\
\ame \lb1-\lb\frac{dX_{\rm B}}{dT_{\rm B}}\rb^2\rb dT_{\rm B}^2\nne
\ame \lb1-V(T_{\rm B})^2\rb dT_{\rm B}^2
\eea
The velocity of B as a function of its coordinate time $T_{\rm B}$ (as measured by A), will be given by Eqn.\eqref{uaccv}, as
\beq
V(T_{\rm B})= \frac{gT_{\rm B}}{\left(1+(gT_{\rm B})^{2}\right)^{\frac{1}{2}}}
\eeq
thus
\begin{IEEEeqnarray}{lrCl}
 \mathrm{d}\tau_{\rm B}({\rm A}) & = & \lb1-\frac{(gT_{\rm B})^{2}}{1+(gT_{\rm B})^2}\rb^{\frac{1}{2}}\mathrm{d}T_{\rm B} \nonumber\\
\ame\frac{1}{ \lb1+(gT_{\rm B})^2\rb^{\frac{1}{2}}}\mathrm{d}T_{\rm B} .
\end{IEEEeqnarray}
This integrates easily as
\beq
\tau_{\rm B}({\rm A})  =  \frac{1}{g}\operatorname{arcsinh}(gT_{\rm B})\label{eq7}
\eeq
satisfying the boundary condition that $\left.\tau_{\rm B}({\rm A}) \right|_{T_{\rm B}=0}=0$.   Therefore for the accelerating period $\alpha$, which goes from $T_{\rm B}: 0\to T'$ (and by symmetry for each accelerating period) we have the elapsed proper time according to A is
\beq
\tau_{{\rm B},\alpha}({\rm A})= \frac{1}{g}\operatorname{arcsinh}(gT').\label{34}
\eeq
From Eqns. \eqref{7} and (\ref{eq4}), (\ref{34}) we conclude that the total elapsed proper time of the trip for B according to A is
\begin{equation}
\Delta\tau_{\rm B}({\rm A}) = \frac{2T_0}{\gamma_V} + \frac{4}{g}\operatorname{arcsinh}(gT')\,. \label{eq11}
\end{equation}
We will show below that this is less than $\Delta\tau_{\rm A}({\rm A})=T_F$, the elapsed proper time of A according to A, and hence the travelling twin, B, is younger.

\begin{figure}
\begin{center}
\includegraphics[scale=0.5]{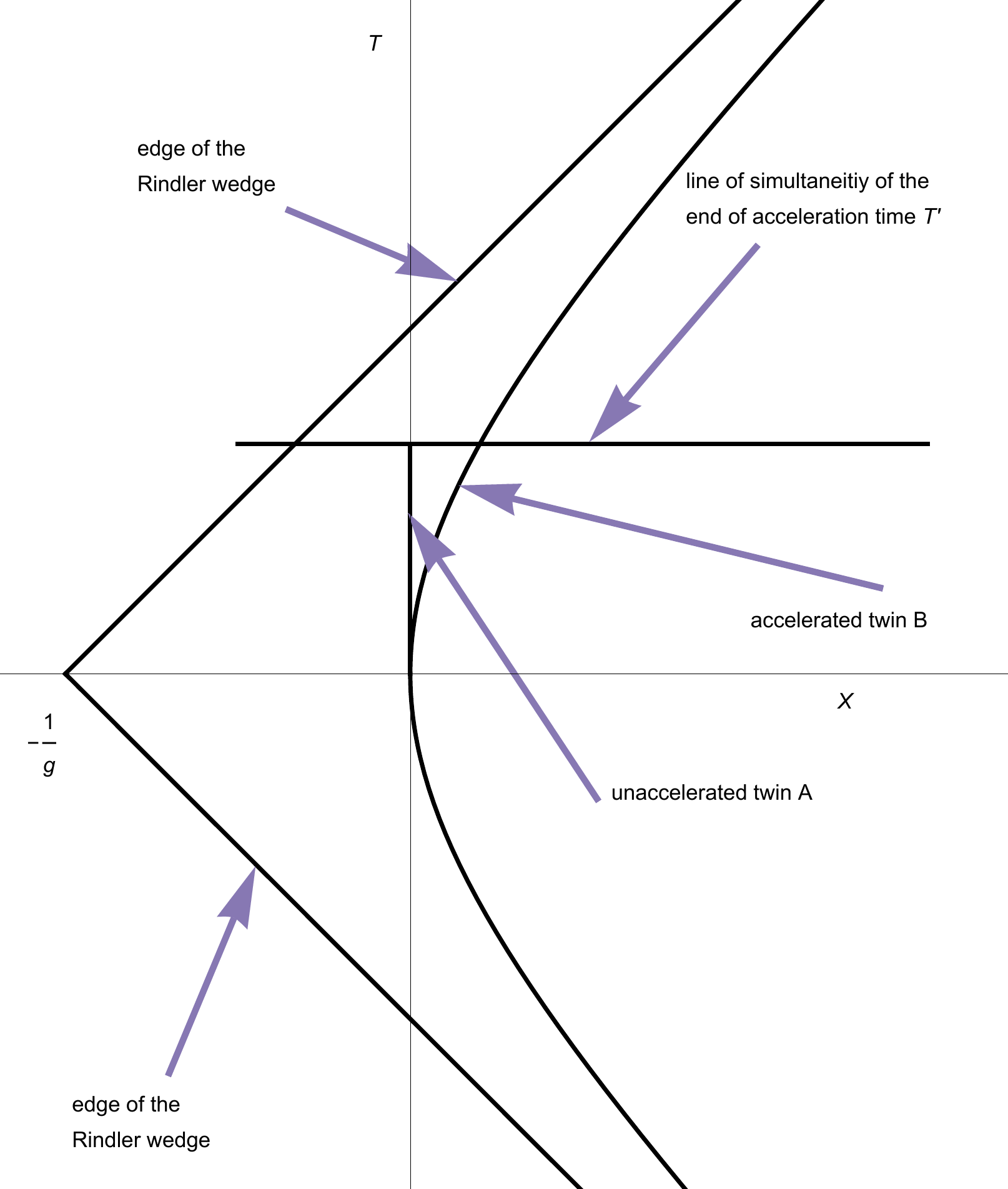}
\caption{Trajectories of A and B for the first accelerating period in the coordinate system of twin A.}\label{fig5}
\end{center}
\end{figure}

We will find it useful in the sequel, to treat the proper time of B, $\tau_{\rm B}$ as the independent variable, inverting Eqn.\eqref{eq7} as
\beq
T_{\rm B}(\tau_{\rm B})  =  \frac{1}{g}\operatorname{sinh}(g\tau_{\rm B}).
\eeq
With Eqn. (\ref{eq28}) and (\ref{eq7}), one can easily determine $X_{\rm B}$ as a function of $\tau_{\rm B}$. Indeed, 
\bea
\lb\frac{dX_{\rm B}(\tau_{\rm B})}{d\tau_{\rm B}}\rb^2\ame\lb\frac{dT_{\rm B}(\tau_{\rm B})}{d\tau_{\rm B}}\rb^2-1\nne
\ame \cosh^2(g\tau_{\rm B}) -1\nne
\ame\sinh^2(g\tau_{\rm B}).
\eea
This integrates trivially as
\beq
X_{\rm B}(\tau_{\rm B})=\frac{1}{g}\cosh(g\tau_{\rm B}) - \frac{1}{g}
\eeq
imposing the boundary condition that $X_{\rm B}(\tau_{\rm B}=0)=0$.  Thus we find a unified expression, which will be used later,
\begin{IEEEeqnarray}{rCl}
X_{\rm B}(\tau_{\rm B}) & = & \frac{1}{g}\cosh(g\tau_{\rm B}) - \frac{1}{g}\label{eq8}\\
T_{\rm B}(\tau_{\rm B}) & = & \frac{1}{g}\sinh(g\tau_{\rm B})\,.\label{eq9}
\end{IEEEeqnarray}
The graphical depiction of the motion for the first acceleration period is given in Figure \eqref{fig5}.

\subsubsection{Resolution of the twin paradox according to A}
The elapsed proper time for A according to A was found to be, Eqn.\eqref{7}
\beq
\Delta\tau_{\rm A}({\rm A}) =  T_F=2T_0 + 4T' \label{45}
\eeq
and the corresponding elapsed proper time of B according to A, from Eqn.\eqref{eq11}
\begin{equation}
\Delta\tau_{\rm B}({\rm A}) = \frac{2T_0}{\gamma_V} + \frac{4}{g}\operatorname{arcsinh}(gT')\,.\label{46}
\end{equation}
The first terms in both Eqn.\eqref{45} and Eqn.\eqref{46} are just the special relativistic elapsed proper times  for the two coasting periods, for each twin.  Since $\gamma_V\ge 1$ clearly the lapse of proper time for A is longer than for B in those two periods,
\beq
2T_0\ge\frac{2T_0}{\gamma_V}.
\eeq

This inequality corresponds to the usual resolution of the twin paradox, when the accelerating periods are neglected, according to the twin A, who does not move.  It is also the source of the twin paradox, if we neglect the accelerated parts of the trajectory and invoke complete symmetry between the coasting portions of the trajectory.  Then, twin B could make the same conclusion about the ``motion'' of twin A, and come up with the paradox that it must be in fact twin A who is younger.  As we will explicitly see, it is wrong to neglect the accelerated part of the trajectory of twin B and it is wrong to imagine that the motion is completely symmetric even for the unaccelerated, coasting parts of the trajectories. 

Looking in detail at the lapse of proper time during the accelerating periods, we can use the explicit analytic formula for the $\operatorname{arcsinh}x=\ln (x+\sqrt{x^2+1})$ to write
\bea
\frac{4}{g}\operatorname{arcsinh}(gT')&\le& 4T'\nonumber\\
\Rightarrow gT'+\sqrt{(gT')^2+1}&\le& e^{gT'}\label{ineq1}
\eea
comparing the lapse of proper time during the accelerating periods. The inequality \eqref{ineq1} is not obviously valid,  however it is easy to prove, see  \footnote{Isolating the square root and then squaring gives
$
(gT')^2+1\le e^{2gT'}-2e^{gT'}gT'+(gT')^2
$
which gives
$
1\le e^{gT'}\lb e^{gT'}-2gT'\rb .
$
This inequality, and hence the original inequality Eqn.\eqref{ineq1}, is valid can be confirmed observing that the RHS is an increasing function of $gT'$ and is equal to 1 at $gT'=0$.  Letting
$
f(gT')=e^{gT'}\lb e^{gT'}-2gT'\rb 
$
and differentiating once with respect to $gT'$ gives 
$
f'(gT')=2e^{gT'}\lb e^{gT'}-(gT'+1)\rb.
$
The last factor $\lb e^{gT'}-(gT'+1)\rb$ is a function of $gT'$ that vanishes at $gT'=0$ and then is a strictly increasing function (This can also be seen by taking its derivative, and observing that the derivative is positive semi definite).  Thus $f'(gT')$ vanishes at  $gT'=0$, and afterwards is strictly positive.  Then, it is easy to see that $f(gT')$ must have a minimum at $gT'=0$ and afterwards is an increasing function.  But $f(0)=1$.  Therefore the inequality, $1\le e^{gT'}\lb e^{gT'}-2gT'\rb $ is satisfied.}.    Therefore, for the accelerating part of the trajectory, we also have
\beq
 \frac{4}{g}\operatorname{arcsinh}(gT')\le 4T'
\eeq
and hence the elapsed proper time during the accelerated part of the trajectory is also greater for twin A than for twin B.

Thus the calculation from A's side clearly gives the expected result, that A, who does not travel, will be older than B,  when B, who does travel, returns.  
\subsection{Interlude}
In this section, we will analyze the reference frame, more generally  the coordinate system, that is appropriate for twin B.  We will impose  that in this coordinate system, twin B's position is always at rest at its origin; twin B does not move in his or her coordinate system.  Then, it cannot be a simple inertial reference frame, as twin B suffers acceleration.  During the coasting periods the coordinate system of twin B will simply be an inertial reference frame, however, during the accelerating periods, it must be something different.  
\subsubsection{Accelerating phase}
\begin{figure}
\begin{center}
\includegraphics[scale=0.6]{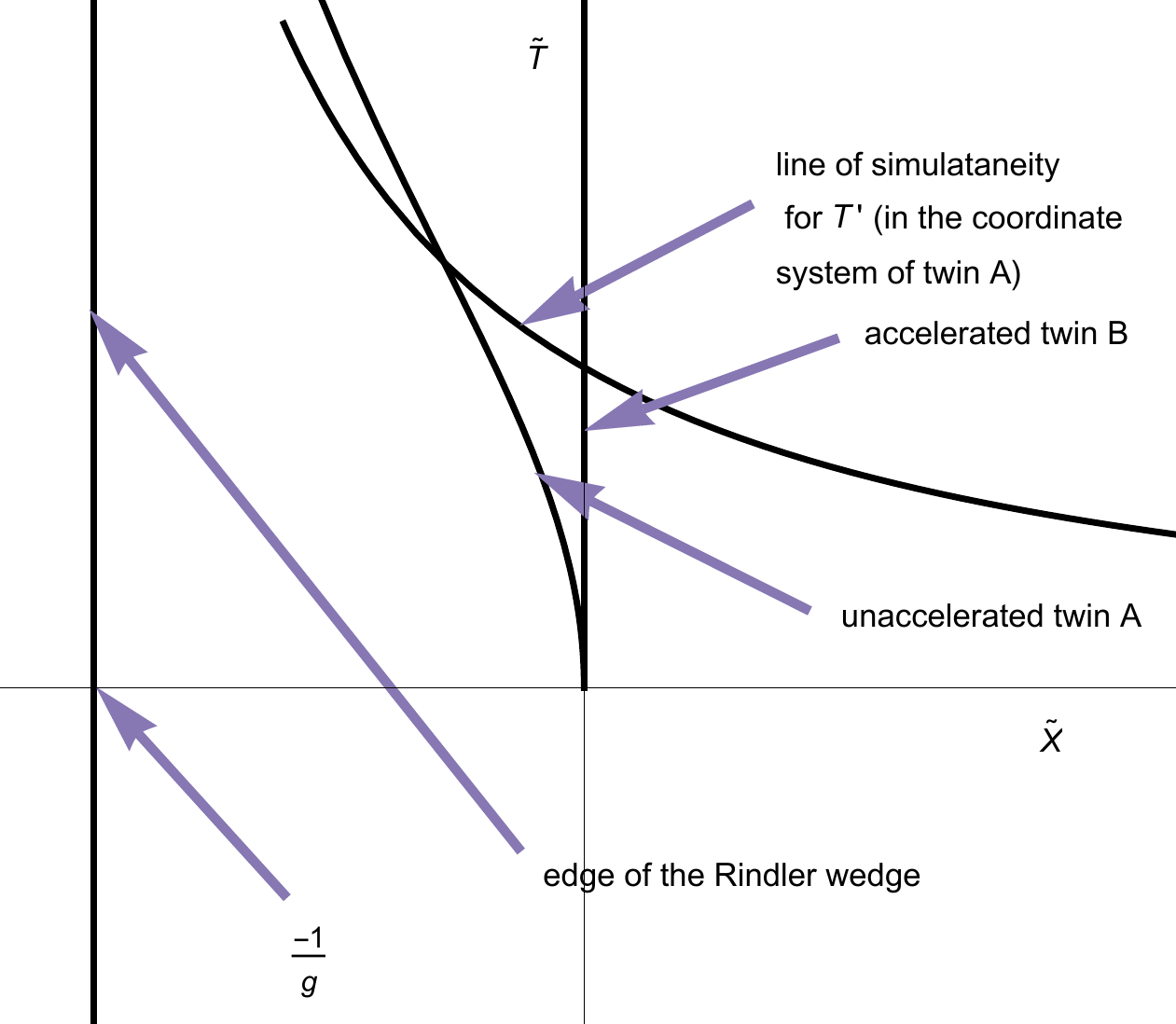}
\caption{Trajectories of A and B for the first accelerating period in the coordinate system of twin B.}\label{fig7}
\end{center}
\end{figure}

The coordinate system during the accelerating phase cannot be an inertial reference frame, as any inertial observer will see twin B as accelerating.  The remainder of this subsection makes precise the trajectories that are depicted in Fig.\eqref{fig7} which is the the new, non-inertial coordinate system that is required.  
Taking a hint from the expressions for the coordinates of B according to A as a function of the proper time of B, Eqns.(\ref{eq8},\ref{eq9}), we consider the transformation of coordinates between $(T,X)$ for twin A and $(\tilde T, \tilde X)$ for twin B
\begin{IEEEeqnarray}{rCl}
X& = & \left(\frac{1}{g} + \tilde X \right)\cosh(g\tilde T) - \frac{1}{g} \label{eq12aa} \\
T & = & \left(\frac{1}{g} + \tilde X \right)\sinh(g\tilde T) \label{eq13bb}\,.
\end{IEEEeqnarray}
with the corresponding (somewhat more complicated) inverse transformation
\bea
\tilde X\ame\sqrt{\lb\lb X+\frac{1}{g}\rb^2-T^2\rb}-\frac{1}{g}\\
\tilde T\ame\frac{1}{g}\operatorname{arctanh}\lb\frac{T}{X+\frac{1}{g}}\rb .
\eea
These coordinates, $(\tilde T,\tilde X)$, are called Kottler-Rindler \cite{Kottler1914,Rindler1960,Rindler1966} coordinates and they are valid in the Kottler-Rindler wedge defined by $X\in[ -\frac{1}{g},\infty]$ and $|T|\le X+\frac{1}{g}$ or correspondingly, $\tilde X\in[-\frac{1}{g},\infty]$ and $\tilde T\in[-\infty,\infty]$.  There are in fact, a multitude of coordinates that can be assigned to twin B, however the Kottler-Rindler system is convenient since there is a physical interpretation for $(\tilde T,\tilde X)$.  Clearly for $\tilde X=0$ we regain Eqns. \eqref{eq8} and \eqref{eq9}, adding labels to make clear  these are the positions and time of B according to A in terms of $\tilde T$ according to B,
\begin{IEEEeqnarray}{rCl}
X_{\rm B}& = & \frac{1}{g} \cosh(g\tilde T_{\rm B}) - \frac{1}{g} \label{eq12aa0} \\
T _{\rm B}& = & \frac{1}{g} \sinh(g\tilde T_{\rm B}) \label{eq13bb0}\,.
\end{IEEEeqnarray}
Thus we can confidently ascribe the coordinates $(\tilde T,\tilde X)$ to twin B, and in these coordinates, twin B sits at $\tilde X=0$ throughout the accelerating phase.  This also means that the coordinate $\tilde T$ can in fact be identified with the proper time for twin B by comparison again with Eqns. \eqref{eq8} and \eqref{eq9}.  We will verify this fact by explicitly computing the metric in the coordinates $(\tilde T,\tilde X)$.  

Different constant values of $\tilde X$ correspond to objects or observers that are moving with respect to twin A, but staying at a fixed coordinate distance $\tilde X$ from twin B during the accelerating phase.  At fixed $\tilde X$, Eqns, \eqref{eq12aa} and \eqref{eq13bb} correspond to hyperbolas in the coordinate system $(X,T)$.  Indeed
\beq
\lb X+\frac{1}{g}\rb^2-T^2=\left(\frac{1}{g} + \tilde X \right)^2\label{hyperbola}
\eeq
which is a family of hyperbolas parametrized by $\tilde X$.  Thus an observer with fixed $\tilde X$ moves along the corresponding hyperbola as $\tilde T$ evolves.  This observer is also a uniformly accelerated observer however with acceleration $g/\lb 1+g\tilde X\rb$, with the initial position that at $\tilde T=0$ this observer is at the position $(T,X)=(0,\tilde X)$.  The coordinate distance according to twin A, between an observer at $\tilde X=0$ and one at fixed $\tilde X \ne 0$ is not constant.  This distance grows proportional to $\cosh(g\tilde T)$.  Hence, for example, a rigid body must have different accelerations from one end to the other if it is to not contract or expand during its trajectory.   We will not pursue this aspect of the Kottler-Rindler system of coordinates here.  On the other hand, curves of fixed $\tilde T$ correspond to
\beq
{T}={\lb X+\frac{1}{g}\rb}\tanh (g\tilde T)\label{tanh}
\eeq
which are straight lines of slope ${\tanh( g\tilde T)}$ passing through the point $(-\frac{1}{g} ,0)$, which is the focus of the hyperbolas Eqn.\eqref{hyperbola}.  

The velocity of twin B according to twin A is  given by 
\beq
V_{\rm B}=\frac{dX_{\rm B}}{dT_{\rm B}}=\left.\frac{dX}{d T}\right|_{\tilde X=0}.
\eeq
The constraint $\tilde X=0$ implicitly defines $X$ as a function of $T$ through the hyperbola in Eqn.\eqref{hyperbola}.  When $\tilde X=0$, we are explicitly describing the trajectory of B in coordinates of A, and thus we should replace $X\to X_{\rm B}$ and $T\to T_{\rm B}$.   Differentiating Eqn.\eqref{hyperbola} gives
\beq
\lb X_{\rm B}+\frac{1}{g}\rb\lb\left.\frac{dX_{\rm B}}{d T_{\rm B}}\right|_{\tilde X=0}\rb-T_{\rm B}=0
\eeq
and solving for the velocity gives
\bea
V_{\rm B}\ame\left.\frac{dX_{\rm B}}{d T_{\rm B}}\right|_{\tilde X=0}=\frac{T_{\rm B}}{\lb X_{\rm B}+\frac{1}{g}\rb}=\frac{T_{\rm B}}{\sqrt{\left.\left(\frac{1}{g} + \tilde X \right)^2\right|_{\tilde X=0}+T_{\rm B}^2}}\nne
\ame\frac{gT_{\rm B}}{\sqrt{1+(gT_{\rm B})^2}}\label{velocity}
\eea
as we found previously for the uniformly accelerated observer in Eqn.\eqref{uaccv}.    The accelerating phase of twin B must end when twin A measures time $T'$ in reference frame $R$, as in Fig. \eqref{fig1}.  Hence we define $\tilde T'$, from Eqn.\eqref{eq13bb0} as
\beq
T'=\frac{1}{g} \sinh(g\tilde T')\label{ttp}
\eeq
the coordinate time according to twin B when the acceleration stops, and
\beq
V=\frac{gT'}{\sqrt{1+(gT')^2}}=\tanh(g\tilde T').
\eeq

We can ask where does twin A appear in the coordinates $(\tilde T,\tilde X)$ as far a twin B is concerned?  From Eqn.\eqref{eq12aa} we have the position $X=0$ corresponds to
\beq
0= \left(\frac{1}{g} + \tilde X \right)\cosh(g\tilde T) - \frac{1}{g}.
\eeq
This is easily solved for $\tilde X$ as
\beq
\tilde X(\tilde T)=\frac{1}{g}\lb\operatorname{sech} (g\tilde T)-1\rb
\eeq
 which asymptotes to $\tilde X=-\frac{1}{g}$.   Notice that $\operatorname{sech}(x)\le 1$, hence twin B sees twin A as moving in the negative $\tilde X$ direction, but as far as twin B is concerned, twin A moves in the negative $\tilde X$ direction but never manages to escape, achieving  asymptotically $-\frac{1}{g}$ as $\tilde T\to\infty$, if indeed twin B accelerated forever.  
 
 This boundary is  called the horizon or edge of the Rindler wedge, the accelerated observer cannot see the whole of the Minkowski spacetime.  As we noted before, the limits to the Kottler-Rindler coordinates correspond to the lines, in Minkowski spacetime, 
 \beq
 X=\pm T-\frac{1}{g}.
 \eeq
 These are lines of slope $\pm 1$ that pass through $X=-\frac{1}{g}$ at $T=0$.  
 Of course, in this limit, according to twin A, the position of twin B is at  $X=\frac{1}{g}\cosh (g\tilde T)-\frac{1}{g}\to\infty$ and $T=\frac{1}{g}\sinh(g\tilde T)\to \infty$.  
 
 The point is that the Kottler-Rindler coordinates only cover the patch of Minkowski coordinates with \beq
\lb X+\frac{1}{g}\rb^2\ge T^2
\eeq
which is a wedge of Minkowski space bounded by the lines of slope $\pm 1$ that pass through $X=-\frac{1}{g}$ at $T=0$.  Twin A, who sits at $X=0$ can only move from $T=0$ to $T=\frac{1}{g}$ as $\tilde T\to+\infty$ as is clear from Eqn.\eqref{tanh}.  Of course, this is not a physical restriction, only an artefact of the Kottler-Rindler coordinate system, observers at $ X=0$ can move well past $g T=1$, simply they are no longer part of the Kottler-Rindler coordinate system. 

Thus the apparent motion of twin A according to twin B, cannot be described fully for the accelerating part of the trajectory if the trajectory requires that $gT\ge 1$.  The Kottler-Rindler coordinates are not expansive enough to cover the whole of Minkowski spacetime, and we must patch on new coordinates to cover the part of the spacetime for $gT\ge 1$.  However, for short enough accelerations, the Kottler-Rindler coordinates are perfectly fine.  This is not a great restriction, physically, staying within the Kottler-Rindler wedge only requires that the velocity achieved by the travelling twin be less than $V_B=1/\sqrt 2$,  which is quite relativistic and obtained by putting $g T=1$ in Eqn.\eqref{velocity}.  We will assume that the acceleration lasts for such amount of time that the Kottler-Rindler coordinates describe the trajectories completely, for each twin.  The formulae that we will finally realize, will be analytic functions of the Minkowski coordinates, and we are confident that the machinery of differential geometry will give the same analytic expressions if in fact we must patch on more coordinates to cover the parts of Minkowski spacetime that are not covered by the Kottler-Rindler coordinate system.

The metric in Kottler-Rindler coordinates is easily obtained from the Minkowski metric.  First we find the differentials
\bea
dT&=&d\tilde X\sinh(g\tilde T) +\left(\frac{1}{g} + \tilde X \right)g\cosh(g\tilde T)\\
dX&=&d\tilde X\cosh(g\tilde T) +\left(\frac{1}{g} + \tilde X \right)g\sinh(g\tilde T)
\eea
and then the metric is given by
\bea
d\tau^2&=&dT^2-dX^2\nonumber\\
&=&\lb d\tilde X\sinh(g\tilde T) +\left(\frac{1}{g} + \tilde X \right)g\cosh(g\tilde T)\rb^2\nonumber\\
&-&\lb d\tilde X\cosh(g\tilde T) +\left(\frac{1}{g} + \tilde X \right)g\sinh(g\tilde T)\rb^2\nonumber\\
&=&\lb1+g\tilde X\rb^2d\tilde T^2-d\tilde X^2.\label{rindlermetric}
\eea
Thus we confirm explicitly, for twin B which corresponds to $\tilde X=0$, we have
\beq
d\tau^2=d\tilde T^2
\eeq
and the lapse of proper time $\Delta\tau_{\rm B}({\rm B})=\Delta\tilde T_{\rm B}$, the lapse of coordinate time of B.  We underline that this is only true for B, for other observers or objects, for example twin A,  which move at $\tilde X\ne 0$ will not have this simple relation between the proper time and the coordinate time, and the full metric given in Eqn.\eqref{rindlermetric} must be used.   

\subsubsection{Coasting phase}
During the coasting phase the coordinate system of twin B will be an inertial reference frame.  Thus we will have a simple Lorentz transformation between the coordinates of A and B:
\bea
X-X'&=&\gamma_V\lb\tilde X+ V\lb\tilde T-\tilde T'\rb\rb\label{ltx}\\
T-T'&=&\gamma_V \lb V\tilde X+\lb\tilde T-\tilde T'\rb\rb\label{ltt}
\eea
The Lorentz transformation has been appropriately shifted so that when the coordinates of B, $(\tilde T,\tilde X)=(\tilde T', 0)$, ($\tilde T'$ was defined by Eqn.\eqref{ttp}), the coordinates A are given by $(T,X)=(T',X')$ which is the start of the coasting period.  

According to A, the coasting period lasts for a coordinate time $T_0$ during which B moves from $X'$ to $X'+L$ at velocity $V$.  From B's point of view, the point with coordinates $X'+L$ according to A, is moving towards B with velocity $-V$ while B is always just staying at $\tilde X=0$.  Replacing $X=X'+L$ and $(\tilde T,\tilde X)=(\tilde T',\tilde L)$ in Eqn.\eqref{ltx} we find
\beq
L=\gamma_V\tilde L
\eeq 
and therefore, $\tilde L=L/\gamma_V<L$.  This is the first clear sign that the two coasting periods are not symmetric.  The distance to the beginning of the decelerating phase according to A is L, but according to B it is $L/\gamma_V$.  Why does this asymmetry come about?    It is because there is a physical turn around point, a distance $L$ away from when B starts to coast, according to A.   A measures this distance at rest, and therefore $L$ represents the proper length of the space interval between A and whatever the turn around point is.  There is a physical difference between the two twins, A just stays put and is always in an inertial reference frame in which the turn around point is also at rest.  B moves and we know that B moves because he or she suffers acceleration.   This same turn around point approaches B at velocity $-V$ but is only a distance $L/\gamma_V$ away, according to B.  

This asymmetry of the coasting periods does not seem to have been clearly identified in the extant literature.  It is this asymmetry that can be used to resolve the twin paradox if the calculation is done according to twin A.  However, if the calculation is done according to twin B, this asymmetry only exacerbates the twin paradox.  If only this asymmetry is taken into account, then B finds that A should be even younger than what A finds B to be.    In the calculation according to B, the accelerating periods, especially the acceleration at the turn around point, play a crucial role to resolve the paradox.

In conclusion, the usual manner in which the twin paradox is evinced, that the coasting periods are all that really matter, and that they are symmetric therefore each twin should think that the other is equally younger, is simply not true.
\subsubsection{Decelerating phase}
The periods of deceleration, $\gamma$ and $\delta$,  will be associated to somewhat different hyperbolas, but the motion is quite similar.  We will simply  replace $g\to -g$ in Eqns. \eqref{eq12aa} and \eqref{eq13bb}.  This gives
\begin{IEEEeqnarray}{rCl}
X& = & \left(\tilde X -\frac{1}{g} \right)\cosh(g\tilde T) + \frac{1}{g} \label{eqdaa} \\
T & = & -\left( \tilde X -\frac{1}{g}  \right)\sinh(g\tilde T) \label{eqddbb}\,.
\end{IEEEeqnarray}
with corresponding hyperbola
\beq
\lb X- \frac{1}{g} \rb^2-T^2=\left(\tilde X -\frac{1}{g} \right)^2,
\eeq
which is also a family of hyperbolas parametrized by $\tilde X$.   We  realize that we are interested in the left branch of this set of hyperbolas, which is the decelerating branch.   This requires that $\tilde X\le \frac{1}{g}$ which then imposes that  $X\le \frac{1}{g}$.   Because of this, the time $\tilde T$ and $T$ run in the same direction, in Eqn.\eqref{eqddbb}, $ -\left( \tilde X -\frac{1}{g}  \right)$ is positive.  

Simple deceleration is not enough, we want that the trajectory of the decelerating observer to be defined by $\tilde X=0$, to pass through the turn around point at the correct spacetime  point in the coordinates of A, $(T_D,X_D)=(T_0+2T',X_0+2X')$ when $\tilde T=\tilde T_D$ (the value of $\tilde T_D$ is actually not required for our analysis, it can of course be determined as the sum of the lapse of coordinate time of B for the accelerating phase, the coasting phase and the decelerating phase).
Thus we shift the hyperbola as
\begin{IEEEeqnarray}{rCl}
X-X_D& = & \left(\tilde X -\frac{1}{g} \right)\cosh(g\lb\tilde T-\tilde T_D\rb) + \frac{1}{g} \label{eqdaas} \\
T -T_D& = & -\left( \tilde X -\frac{1}{g}  \right)\sinh(g\lb\tilde T-\tilde T_D\rb) \label{eqddbbs}\, .
\end{IEEEeqnarray}
This is now a hyperbola that passes through $X=X_D+\tilde X=X_0+2X'+\tilde X$ and and $T=T_D=T_0+2T'$ when $\tilde T=\tilde T_D$.  Clearly
\beq
\lb X-X_D-\frac{1}{g}\rb^2-\lb T -T_D\rb^2= \left(\tilde X -\frac{1}{g} \right)^2
\eeq
which is the equation of a family of hyperbolas, parametrized by $\tilde X$, symmetric about $\lb T_D,X_D+\frac{1}{g}\rb$.  We choose the left branch (decelerating) by imposing that $\tilde X\le \frac{1}{g}$.  As mentioned above, the hyperbola for $\tilde X=0$ corresponds to the trajectory of twin B and at $\tilde T=\tilde T_D$, twin B will be at the turn around point $(X_D,T_D)$ in the coordinate system of twin A. 

It is easy to check that the metric in these coordinates for the decelerating phase is exactly as before except $g\to -g$
\beq
d\tau^2=\lb 1-g\tilde X\rb^2d\tilde T^2 -d\tilde X^2.
\eeq
To compute the elapsed proper time for twin B during the decelerating phase, we should integrate $\tilde T$ from $\tilde T_D-\tilde T'\to \tilde T_D$ where $\tilde T'$ is defined as the lapse of coordinate time of B during the accelerating phase, during which the coordinate time of A increases from 0 to $T'$.  Then from Eqn.\eqref{ttp}, we have
\beq
\tilde T'=\frac{1}{g}\operatorname{arcsinh}\lb gT'\rb .\label{ptd}
\eeq
It is clear that to decelerate from $V$ to 0 will take the same amount of coordinate time for B as to accelerate from 0 to $V$.    As the position of B is at $\tilde X=0$, the lapse of coordinate  time of B according to B during the decelerating phase is simply equal to the lapse of proper time of B according to B and hence $\Delta \tau_{{\rm B},\gamma} ({\rm B} )=\tilde T'=\frac{1}{g}\operatorname{arcsinh}\lb gT'\rb $.
\subsection{Elapsed proper time of B according to B}
To compute the elapsed proper time of B according to B is very easy since according to B, B stays put at $\tilde X=0$ for the entire journey.  For accelerating and decelerating parts of the trajectory $\alpha$ and for $\gamma$, we have already calculated the elapsed proper time, from Eqn.\eqref{ttp} and from Eqn.\eqref{ptd} we have
\beq
\Delta\tau_{{\rm B},\alpha}=\Delta\tau_{{\rm B},\gamma}=\tilde T'=\frac{1}{g}\operatorname{arcsinh}\lb gT'\rb .\label{ptbag}
\eeq
For the coasting period $\beta$ we must compute the change in the coordinate time of B according to B and then convert this to proper time of B according to B.   The coasting period starts at $(\tilde T,\tilde X)=(\tilde T',0)$ and ends when the coordinates of A are $(T'+T_0,X'+L)$.  Replacing the end coordinates into Eqn.\eqref{ltt}, imposing $\tilde X=0$ and labelling $\tilde T_0+\tilde T'$ as the  coordinate time when the coasting phase stops,  (clearly $\tilde T_0$ is the elapsed coordinate time during the coasting phase according to B),  we find 
\beq
T_0=\gamma_V\lb (\tilde T_0+\tilde T')-\tilde T'\rb ,
\eeq
which gives
\beq
\tilde T_0=\frac{T_0}{\gamma_V}.\label{ebtcp}
\eeq
The elapsed coordinate time for B is shorter than the elapsed coordinate time for A which makes plain the lack of symmetry between the two twins during the coasting period.  

The corresponding elapsed proper time of B then is then easily computed as the metric is just the Minkowski metric
\beq
d\tau^2=d\tilde T^2-d\tilde X^2= d\tilde T^2
\eeq
since $d\tilde X=0$.  Integrating from $\tilde T'\to \tilde T'+\tilde T_0$ gives
\beq
\Delta\tau_{{\rm B},\beta}({\rm B})=\tilde T_0=\frac{T_0}{\gamma_V}.\label{ptbb}
\eeq
Adding all the contributions together from Eqns. \eqref{ptbag} and \eqref{ptbb} and invoking symmetry for the return part of the journey, we find
\bea
\Delta\tau_{{\rm B}}({\rm B})&=&\lb \Delta\tau_{{\rm B},\alpha}({\rm B})+\Delta\tau_{{\rm B},\beta}({\rm B})+\Delta\tau_{{\rm B},\gamma}({\rm B})\rb\times 2\nonumber\\
&=&2\frac{T_0}{\gamma_V}+4\frac{1}{g}\operatorname{arcsinh}\lb gT'\rb\label{ptbb}
\eea
which is exactly the same as what A calculated, Eqn.\eqref{46}
\beq
\Delta\tau_{{\rm B}}({\rm B})=\Delta\tau_{{\rm B}}({\rm A}).
\eeq
\subsection{Elapsed proper time of A according to B}
Finally, we must compute the elapsed proper time of A according to B.  This is the most complicated of the calculations.  We have only seen such a calculation using the machinery of differential geometry and general relativity \cite{Perrin}.  Here we will show how to do the calculation using only simple changes of variable, which are actually the coordinates.  No knowledge of general relativity or differential geometry is required.  It is clear that we must only compute the proper time for the phases $\alpha,\beta,\gamma$, the rest of the trajectory just gives twice this answer.  
\subsubsection{Accelerating phase}
During the accelerating phase we have found the coordinate transformation given in Eqns.\eqref{eq12aa} and \eqref{eq13bb} is appropriate.   In this transformation, twin A stays put at $X=0$.  This gives
\begin{IEEEeqnarray}{rCl}
0& = & \left(\frac{1}{g} + \tilde X \right)\cosh(g\tilde T) - \frac{1}{g} \label{eq12AX} \\
T & = & \left(\frac{1}{g} + \tilde X \right)\sinh(g\tilde T) \label{eq13AT}\,
\end{IEEEeqnarray}
which can be solved for $\tilde X$ as
\beq
\tilde X_{\rm A}=\frac{1}{g}\lb\operatorname{sech}(g\tilde T_{\rm A})-1\rb.\label{xt}
\eeq
Replacing this in Eqn.\eqref{eq13AT} gives
\beq
T_{\rm A}=\frac{1}{g} \tanh(g\tilde T_{\rm A}).\label{95}
\eeq
We note that remarkably, this relation between $T_{\rm A}$ and $\tilde T_{\rm A}$ for the motion of A according to B is not the same as the relation between $T_{\rm B}$ and $\tilde T_{\rm B}$ for the motion of B according to B given in Eqn.\eqref{eq13bb0}:
\beq
T_{\rm B}=\frac{1}{g} \sinh(g\tilde T_{\rm B}).\label{tbttb}
\eeq
The accelerating phase terminates when $T_{\rm B}=T'$ the time of B according to A and in principle when $T_{\rm A}=T'$ the time of A according to A.  These times (according to A) give rise to different times for when the acceleration stops, according to B.  This is easily understood by the notion of the relativity of simultaneity.  The two events when $T_{\rm A}=T'$ and when  $T_{\rm B}=T'$ are simultaneous according to A, but occur at different spatial points.  Although they are simultaneous for A, they are not for B.  Drawing the surfaces of simultaneity, in each twin's reference frame,  we would see, for example, that when B's proper time is such that the initial acceleration period stops, two different times are relevant to describe A. In A's frame, B stopping is simultaneous to A having aged $T=T'$, which is the description encoded in Eqn.\eqref{tbttb}.  However, what is now relevant is that in B's frame, B stopping is simultaneous to A having aged according to Eqn.\eqref{95}.

There is also another mismatch that occurs that is worth elaborating.  The velocity of A according to B at the end of the accelerating phase will be
\bea
V_{\rm A}({\rm B})=\left.\frac{d\tilde X_{\rm A}}{d\tilde T}\right|_{\tilde T'}\ame-\tanh (g \tilde T')\operatorname{sech} (g \tilde T')\nne
&\ne& -V=-\tanh(g\tilde T')
\eea
Thus B at the end of the accelerating phase, does not see A receding with velocity $-V$ but with a somewhat smaller velocity.  When B stops accelerating, discontinuously or in reality, rather brusquely, he or she quickly adopts the inertial coordinate system of Eqn.\eqref{ltx} and \eqref{ltt}.  Physically, this change of coordinate system must occur continuously,  however it is normally the case that it occurs rather fast.  Treating it as if it is a discontinuous change of coordinate system, then there is a discontinuous change of the velocity from $V_{\rm A}({\rm B})\to -V$.  It is understood that the 4-velocity of A, $\frac{d\tilde X_{\rm A}^\mu}{d\tau_{\rm A}({\rm B})}$, in the accelerating Kottler-Rindler coordinate system of B just before the acceleration stops and in the inertial Lorentz coordinate system just after the acceleration stops, is related by the standard tensorial relation between 4-vectors in different coordinates systems, to the 4-velocity of A in the coordinate system of A which is always $(1,0,0,0)$ (adding in the $Y$ and $Z$ coordinates).  Thus, any discontinuity is solely due to a discontinuous change of coordinate system.

The metric for the Kottler-Rindler coordinates is given by Eqn.\eqref{rindlermetric}
\beq
d\tau^2=\lb1+g\tilde X\rb^2d\tilde T^2-d\tilde X^2
\eeq
thus using Eqn.\eqref{xt} for $\tilde X_{\rm A}$
\bea
d\tau_{{\rm A},\alpha}({\rm B})&=&\lb \lb1+g\tilde X_{\rm A}\rb^2-\lb\frac{d\tilde X_{\rm A}}{d\tilde T_{\rm A}}\rb^2\rb^{1/2}d\tilde T_{\rm A}\nne
\ame\lb \operatorname{sech}^2(g\tilde T_{\rm A})-\tanh^2(g\tilde T_{\rm A})\operatorname{sech}^2(g\tilde T_{\rm A})\rb^{1/2}d\tilde T_{\rm A}\nne
\ame\operatorname{sech}^2(g\tilde T_{\rm A})d\tilde T_{\rm A}.\label{sc}
\eea
This integrates trivially and gives, using the relation Eqn.\eqref{tbttb},
\bea
\Delta\tau_{{\rm A},\alpha}({\rm B})\ame\int_0^{\tilde T_B}\operatorname{sech}^2(g\tilde T_{\rm A})d\tilde T_{\rm A}=\frac{1}{g}\tanh (g\tilde T_{\rm B})\nne
\ame\frac{1}{g}\frac{\sinh (g\tilde T_{\rm B})}{\cosh (g\tilde T_{\rm B})}=\frac{T'}{\lb1+(gT')^2\rb^{1/2}}.
\eea
Thus the elapsed proper time of A according to B during the accelerating phase $\alpha$ is less than what A would calculate, $T'$.
\subsubsection{Coasting phase}
During the coasting phase, as we have already ascertained, the lapse of coordinate time for twin B is 
given by Eqn.\eqref{ebtcp}
\beq
\tilde T_0=\frac{T_0}{\gamma_V}.
\eeq
During this lapse of coordinate time of B, twin A moves from $\tilde X=-\tilde X'$ to 
\beq
\tilde X=-\tilde X'-\tilde T_0 V.
\eeq
The elapsed proper time is given by integrating
\bea
\Delta\tau_{\rm A}({\rm B})\ame\int_{\tilde T'}^{\tilde T'+\tilde T_0}d\tau_{\rm A}({\rm B})=\int_{\tilde T'}^{\tilde T'+\tilde T_0}\lb1-V^2\rb^{1/2}d\tilde T\nne
\ame \lb1-V^2\rb^{1/2}\tilde T_0=\frac{\tilde T_0}{\gamma_V}.
\eea
Replacing in for $\tilde T_0$ we find
\beq
\Delta\tau_{\rm A}({\rm B})=\frac{ T_0}{\gamma_V^2}.
\eeq
As $\gamma_V>1$ this is like a double whammy.  During the coasting period, the calculation of the proper time of A according to B is even much smaller than the symmetric calculation of the elapsed proper time of B according to A, as given in Eqn. \eqref{eq4}, $\Delta\tau_{\rm B}({\rm A})=\frac{ T_0}{\gamma_V}$.  The decelerating phase must come to the rescue and give us back the true, full lapse of proper time of A.
\subsubsection{Decelerating phase\label{dp}}
Naively, we might think that the lapse of proper time in the accelerating phase and the decelerating phase for A according to B would be equal.  But this is simply not true.  It is the decelerating phase that makes up for all the time lost that seems to be making A younger than B.

Indeed, we compute the proper time in the same manner for the decelerating phase as for the accelerating phase, notice the metric in Eqn.\eqref{rindlermetric} changes with $g\to -g$,
\beq
d\tau=\lb\lb 1-g\tilde X\rb^2-\lb\frac{d\tilde X}{d\tilde T}\rb^2\rb^{1/2} d\tilde T
\eeq
and replacing $X=0$ in Eqn.\eqref{eqdaas} yields $\tilde X_{\rm A}$ 
\beq
\tilde X_{\rm A}=-\lb X_D+\frac{1}{g}\rb \operatorname{sech}\lb g(\tilde T-\tilde T_D)\rb+\frac{1}{g}.
\eeq
Then we get, with a calculation very similar to that done for Eqn.\eqref{sc}
\bea
d\tau_{{\rm A}\gamma}({\rm B})\ame\lb\lb 1+gX_D\rb\operatorname{sech}^2\lb g(\tilde T_{\rm A}-\tilde T_D)\rb\rb d\tilde T_{\rm A}\nne
\eea
and integrating from $\tilde T_D-\tilde T'$ to $\tilde T_D$ gives
\bea
\Delta\tau_{{\rm A}\gamma}({\rm B})\ame\lb 1+gX_D\rb\frac{1}{g}\tanh\lb g\tilde T'\rb\nne
\ame \lb 1+g(X_0+2X')\rb\frac{T'}{\lb1+(gT')^2\rb^{1/2}}.\nne
\eea
We note that $\Delta\tau_{{\rm A}\gamma}({\rm B})$ can be as large as required because of the additional term 
\beq
g(X_0+2X')\frac{T'}{\lb1+(gT')^2\rb^{1/2}}
\eeq 
which compensates for the smaller lapse of proper time of A according to B in the phases $\alpha$ and $\beta$.
\subsubsection{Proper time of {\rm A} as calculated by  {\rm B} }
Now finally we can put all the pieces together to get, since $\Delta\tau_{{\rm A}}({\rm B})=\lb\Delta\tau_{{\rm A}\alpha}({\rm B})+\Delta\tau_{{\rm A}\beta}({\rm B})+\Delta\tau_{{\rm A}\gamma}({\rm B})\rb\times 2$
\bea
\Delta\tau_{{\rm A}}({\rm B})/2\ame\Delta\tau_{{\rm A}\alpha}({\rm B})+\Delta\tau_{{\rm A}\beta}({\rm B})+\Delta\tau_{{\rm A}\gamma}({\rm B})\nne
\ame \frac{T'}{\lb1+(gT')^2\rb^{1/2}}+\frac{ T_0}{\gamma_V^2}\nne
&+& \lb 1+g(X_0+2X')\rb\frac{T'}{\lb1+(gT')^2\rb^{1/2}}.\nne
\eea
Then using 
\beq
\frac{V}{g}=\frac{T'}{\lb1+(gT')^2\rb^{1/2}}
\eeq
\beq
X'=\frac{1}{g}\lb\lb1+\lb gT'\rb^2\rb^{1/2}-1\rb 
\eeq
and $X_0=VT_0$
we get
\bea
\Delta\tau_{{\rm A}}({\rm B})/2\ame\frac{V}{g}+\lb1-V^2\rb T_0+\lb 1+gVT_0\rb \frac{V}{g}\nne
&+&2\frac{V}{g}\lb\lb1+\lb gT'\rb^2\rb^{1/2}-1\rb \nne
\ame T_0+2\lb1+\lb gT'\rb^2\rb^{1/2} \frac{T'}{\lb1+(gT')^2\rb^{1/2}}\nne
\ame T_0+2T'.
\eea
Therefore we reproduce that the elapsed proper time of A according to B is
\beq
\Delta\tau_{{\rm A}}({\rm B})=2T_0+4T'=\Delta\tau_{{\rm A}}({\rm A})
\eeq
and from Eqn.\eqref{ptbb} and Eqn.\eqref{46}, we have
\beq
\Delta\tau_{{\rm B}}({\rm B})=2\frac{T_0}{\gamma_V}+4\frac{1}{g}\operatorname{arcsinh}\lb gT'\rb=\Delta\tau_{{\rm B}}({\rm A})
\eeq
and of course
\beq
\Delta\tau_{{\rm B}}({\rm A})=\Delta\tau_{{\rm B}}({\rm B})<\Delta\tau_{{\rm A}}({\rm B})=\Delta\tau_{{\rm A}}({\rm A}).
\eeq

In conclusion, the twin paradox is completely resolved, the sedentary twin A  is older than the travelling twin B after the journey, and we have explicitly shown how to do the calculation of the elapsed proper time of each twin, according to each twin.

\section{Comments on Maudlin's analysis}
\subsection{Is acceleration crucial to the resolution of the twin paradox?}
\begin{figure}[h]
\begin{center}
\includegraphics[scale=0.4]{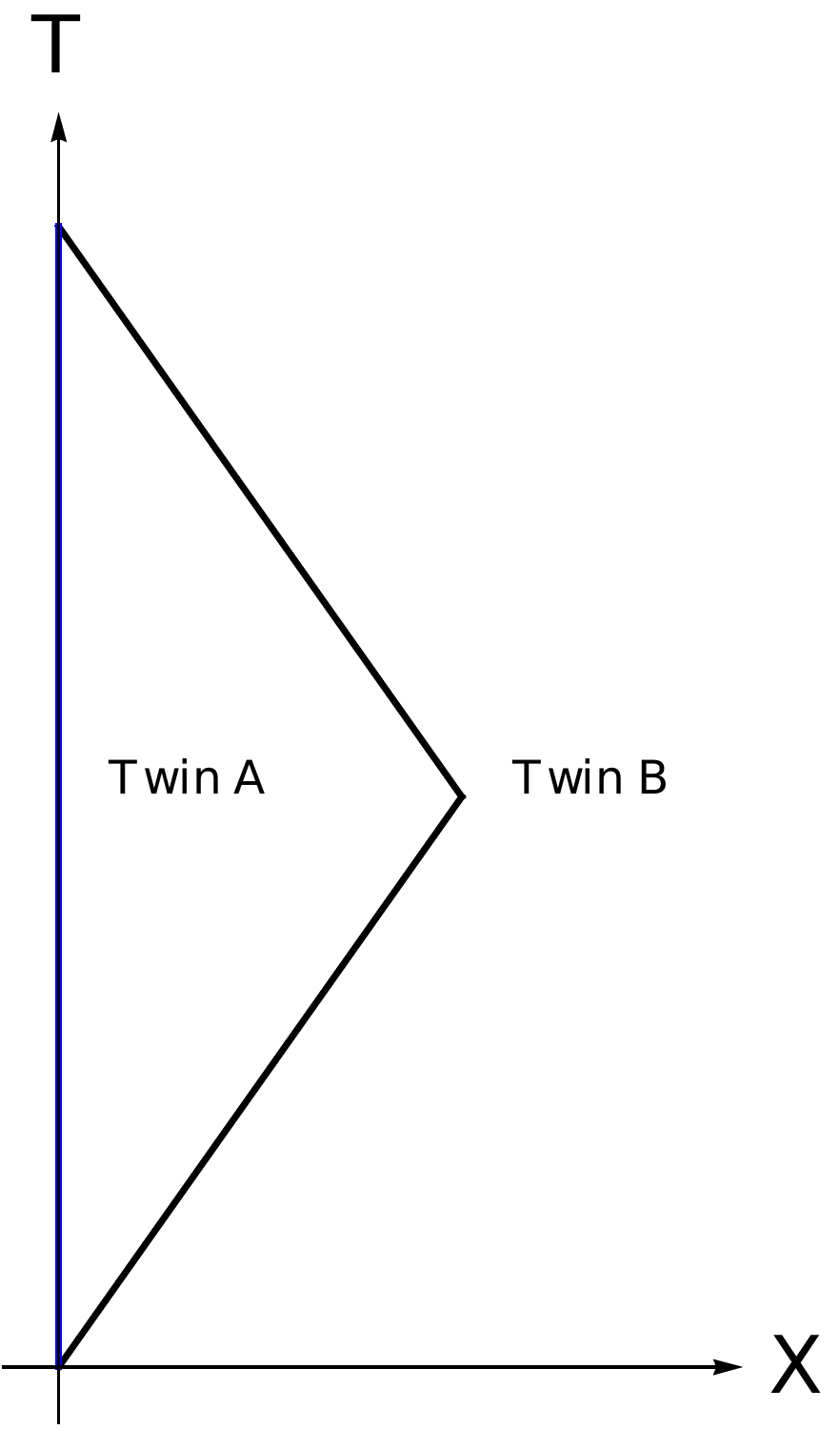}
\caption{Infinite proper acceleration version of the twin paradox}\label{fig2}
\end{center}
\end{figure}

Philosopher of physics T. Maudlin analyzed the  twin paradox in his book \textit{Philosophy of Physics: Space and Time}, \cite{Maudlin2012}.  Maudlin quotes Feynman at length from the book \cite{feynman} and maintains that everything in the explanation found there is wrong.  Feynman is quoted as saying 
\begin{quote}
So the way to state the rule is to say that the man who has felt the accelerations, who has seen things fall against the walls, and so on, is the one who would be the younger; that is the difference between them in an ``absolute'' sense, and it is certainly correct.
\end{quote}
Maudlin proceeds to try to demonstrate that acceleration plays no role in explaining the end result.  In his analysis, he notes that the lengths of the accelerating parts of B's worldline can be made as small as possible, and so he argues that the accelerating periods play no significant role in the resolution of the twin paradox, as depicted in Fig.\eqref{fig2}.   As he says on page 83: ``the issue is how \textit{long} the world-lines are, not how \textit{bent}''.  

Let us first try to understand why considering the accelerating part of B's journey is extremely important in understanding the original twin paradox.    Maudlin's argument is that we can calculate the proper time of each twin in a single Lorentz frame, and that it will yield the correct result, thus nothing more needs to be said.   The same calculation done in any  Lorentz frame will also lead to the same conclusion: twin A must be older than twin B at the end of the journey.   So one may conclude that there is no true paradox, the theory of special relativity does not lead to any contradictory results. 

To get to the crux of the issue we must ask: \textit{Why would anyone think that the situation may be paradoxical?}   Exposure to the theory of special relativity and the notion that all motion is relative, (certainly inertial motion) one would think the situation is paradoxical because of the symmetry between A and B.  A would describe the relative motion of B in ``exactly the same way'' as B would describe the relative motion of A, and consequently both twins should think the other must come out to be younger.   To uncover the fallacy of this argument and resolve the paradox, we have to find a source for the asymmetry of the situation.  In flat infinite Minkowski spacetime, \ie in the original version of the paradox, as Feynman says \cite{feynman}, it is the acceleration that gives rise to the asymmetry.  

Doing the calculation of proper times in B's reference frame and confirming the result obtained when doing the calculation in A's reference frame,  might not be strictly necessary in finding the age of both twins when they meet after the journey, but it is crucial in demonstrating why the twin paradox is not truly a paradox.  From Section \eqref{dp}, we see that during the turnaround period, B associates a time lapse to A which grows very rapidly,  balancing the slower aging of A, from B's point of view, during coasting periods and during the accelerating part of B's trajectory when he or she is near A.  This alone demystifies the sole argument that could lead people to think a paradox exists, \ie that during coasting periods, both twins think the other ages less.  

As we have shown, whether acceleration plays a small role or an important role, depends on who is doing the calculation.   If it is twin A, then it is quite correct to neglect the acceleration (if it is for a short time compared to the coast times as in Fig.\eqref{fig2}), but if it is twin B, then it is completely incorrect to neglect the acceleration. The complete resolution of the paradox is obtained by doing the calculation of the elapsed proper time of each twin, according to each twin.  It is not correct to neglect the acceleration for the twin that takes the journey, and for that twin's calculation, it is crucial to take into account the accelerated parts of the trajectory.   Therefore it is simply wrong to say that acceleration plays no role in the resolution of the twin paradox.  

Is acceleration crucial to the twin paradox?  Consider the situation in flat infinite spacetime, where no twin ever undergoes acceleration, but they are moving at constant speed relative to each other.  Then both twins consider the other to be aging slower, but this is perfectly well understood in special relativity and time dilation.  However, without acceleration, the twins never come back together and so this in no way constitutes a paradox.  To obtain the paradox, one twin must necessarily undergo acceleration.  Period.  No acceleration, no paradox. 

Many different analyses of the resolution of the twin paradox exist, however we feel the one presented here that highlights the crucial role of acceleration, tackles it with the most clarity.   An enlightening discussion of the many existing analyses can be found in \cite{Baez}.
\subsection{Modifying the twin paradox}
Maudlin further elaborates his point by stating that twin A could in fact accelerate the same amount as twin B or even more than twin B,  and still be older, as he illustrates through the following example.  

\begin{figure}[h]
\begin{center}
\includegraphics[scale=0.4]{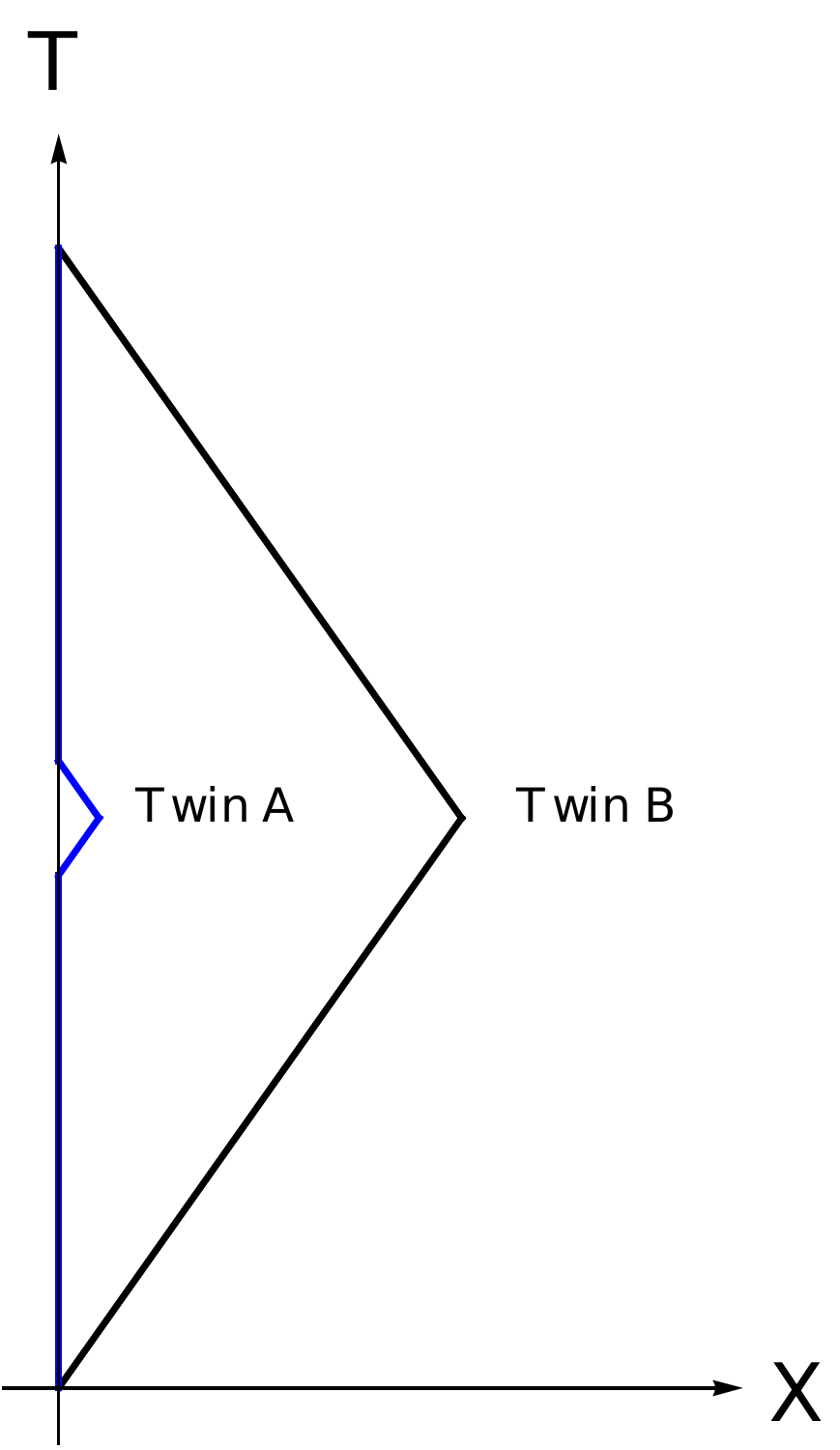}
\caption{New version of the twin paradox}\label{fig3}
\end{center}
\end{figure}

Maudlin modifies the circumstances of the twin paradox by considering a situation, where twin A undergoes one small period of accelerated motion at the middle of the journey, characterized by a triangular ``bump'' on the Minkowski diagram of the frame of the original twin A (see Figure \eqref{fig3}).  
It is clear that the acceleration of A can be equal to or even greater than the acceleration of B.  However, because A's worldline is still longer than B's,  A ends up older than B, even though A underwent an equal or greater amount of acceleration.   Maudlin's conclusion then is that acceleration plays no role in the resolution of the twin paradox.  

But where is the paradox in this case?  The motion of the two twins is in no sense symmetric.  There is no symmetry argument that one could make that would lead to a paradox.  Simply we would have to do the calculation of the elapsed proper time of each twin according to each twin and find the obvious fact  that twin B ends up younger.  This modification of the circumstances does not in the slightest take away from the fact that to do the calculation of the elapsed proper time according to each twin in this case, we must take great care to calculate what happens during the accelerating phases.  

It is true that when analyzing the problem from the frame of reference of twin A, one can  neglect the \textit{length} of the accelerating part of the travelling twin's worldline, when it is much shorter than the coasting part (as in Fig\eqref{fig2}), and one does of course find the travelling twin B to be younger.  But the \textit{presence} of acceleration is necessary in even coming up with a paradox at all.   This is why introducing multiple periods of accelerations is specious, it has no bearing on the original paradox.   

Thus for the situation described by Maudlin as in Fig\eqref{fig3}, we have to ask ourselves: ``Would there be any conceptual benefit or clarification in analyzing the modified situation in any non-inertial reference frame, specifically the frames of reference of twin A or twin B, as both undergo periods of acceleration?''  We know we could do it, and in the end we would (and must) obtain the same result as having analyzed it in a convenient Lorentz frame, for example that Lorentz frame in which twin A is initially at rest.  But would it provide any new insights into the understanding of the original twin paradox?  This is the true question one has to ask when trying to gain understanding of a paradox that doesn't reside in the calculations \textit{per se}, but is paradoxical due to a faulty understanding of the underlying physical circumstances.  Coming up with a situation which no one would think is paradoxical does not help in resolving the original paradox, it only clouds the analysis.  

Therefore we do not concur that this example has any bearing on whether or not acceleration is crucial  to the resolution of the original twin paradox.    It is simply clear that viewing the original paradox in B's frame, requires analyzing the accelerating periods and this analysis provides a powerful conceptual understanding of the aging process of both twins, which would otherwise not be explored if we had restricted ourselves to computing the results only in the reference frame of A.  

Maudlin imagines other modifications of the circumstances of the twin paradox, which we will briefly describe below, but which to us do not add clarity to the issue.  On pages 82-83 of his book, he states:
\begin{quote}
In Minkowski spacetime, at least one of the twins must accelerate if they are to get back together: as mentioned above, a pair of straight lines in Minkowski spacetime can meet at most once. This is incidental to the effect: in General Relativity, twins who are both on \textit{inertial} trajectories at all times can meet more than once, and show differential aging when they meet.\cite{Maudlin2012}
\end{quote}
So then Maudlin considers other versions of the twin paradox,  in curved spacetime or closed, flat spacetime.  Maudlin mentions these new situations again in an attempt to refute the relevance of the acceleration in the original twin paradox.  However, although these new situations constitute perhaps new paradoxes, they have no bearing on the original twin paradox.  When we consider, for example, the twin paradox in a spacetime with a closed spatial loop, there has to be something other than acceleration causing the asymmetry in the situation, so that the twin in the spaceship who ``travels the whole universe'' and comes back is younger than his brother.  In that case, the source of the asymmetry resides in the non-trivial topology of the considered spacetime (for a detailed analysis, see \cite{Spacetopology,CompactSpaces}).  In these new situations, global Lorentz symmetry is broken and some observers are preferred for maximal aging along inertial trajectories.  However this case in no way has any bearing on the understanding that the source of asymmetry in the original twin paradox is the acceleration.  The new paradoxes have nothing to say about the original one.  The two problems are simply different, and so it is perfectly consistent that both are explained by different mechanisms, the explanation in one case has no application in the other.  

\section{Conclusion}
An analysis of the twin paradox from the point of view of both twins was performed. It was found that during the turnaround period, the travelling twin B, associates to the sedentary twin A, a lapse of proper time which goes by just fast enough so that it accounts for the A's apparent slower aging (according to B) during coasting period and the initial accelerating period.  It was then argued that this particular analysis provided interesting insights to the problem, namely in that it helped describe how exactly the asymmetry arises between both twins' journeys \ie how the periods of acceleration affected  twin B's analysis of events.  It was also argued that one should be careful in invoking different versions of the twin paradox to explain a point about the ``original'' version: in the end, each version is a different problem, and the explanation of one obviously cannot carry to another.

\section{Acknowledgements}

This work was supported by the NSERC of Canada, B.S. specifically through an Undergraduate Student Research Award (USRA) and J. G. thanks Dicyt-USACH.  We thank Louis-André Hamel and Richard MacKenzie for an in depth reading of the manuscript and for making useful comments.  We thank the Inter-University Center for Astronomy and Astrophysics, Pune, India, the Indian Institute of Science, Education and Research Pune, Pune, India, the Departamento de Fisica, Universidad de Santiago, Santiago, Chile and the Bahamas Advanced Study Institute and Conferences (BASIC), Stella Maris, Long Island, Bahamas for hospitality, where some of this work was written up.


\bibliographystyle{apsrev}
\bibliography{TwinParadoxFinal}

\end{document}